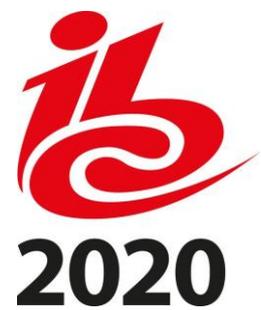

# MPEG MEDIA ENABLERS FOR RICHER XR EXPERIENCES

E. Thomas[1], E. Potetsianakis[1], T. Stockhammer[2], I. Bouazizi[3],
M-L. Champel[4]

[1]TNO, The Netherlands,
[2]Qualcomm Incorporated, Germany,
[3]Qualcomm Incorporated, USA,
[4]Xiaomi, China

## ABSTRACT

With the advent of immersive media applications, the requirements for the representation and the consumption of such content has dramatically increased. The ever-increasing size of the media asset combined with the stringent motion-to-photon latency requirement makes the equation of a high quality of experience for XR streaming services difficult to solve. The MPEG-I standards aim at facilitating the wide deployment of immersive applications. This paper describes part 13, Video Decoding Interface, and part 14, Scene Description for MPEG Media of MPEG-I which address decoder management and the virtual scene composition, respectively. These new parts intend to make complex media rendering operations and hardware resources management hidden from the application, hence lowering the barrier for XR application to become mainstream and accessible to XR experience developers and designers. Both parts are expected to be published by ISO at the end of 2021.

## INTRODUCTION

Extended Reality (XR) is an umbrella term for immersive experiences that includes Virtual Reality (VR), Mixed Reality (MR) and Augmented Reality (AR). These applications utilize computationally demanding technologies such as 360-degree video, spatial audio, 3D graphics, etc. In order to successfully combine these media technologies in power constrained end-devices, strict synchronization of the representations and resource allocation is paramount. Currently, this is achieved with tailored application-specific solutions. In many cases, application developers do not have access to advanced hardware resources, especially when it comes to real-time media in immersive experiences. In order to enable interoperability and efficient usage of device resources, the Moving Picture Experts Group (MPEG) is working on part 13 and part 14 of the ISO/IEC 23090 MPEG Immersive (MPEG-I) standard, both expected to be published by ISO at the end of 2021.

MPEG-I consists of several parts, addressing different needs of XR systems. This paper focuses thus on part 13, Video Decoding Interface (VDI) and part 14, Scene Description for Immersive Media (SD).

VDI addresses the need for managing simultaneous decoding of video elementary streams in an efficient and synchronised fashion. Simultaneous decoding is necessary due to the large number of video and media assets required for immersive experiences (e.g. 360-degree video, video-based point cloud objects, 2D video textures, etc.). In addition, some of those assets may be encoded as a set of independent subparts of the content (e.g. tiles in 360 video tiled-streaming, components in video-based point clouds), and thus have to be decoded separately and then combined together to obtain the final output of the decoding process, which in turns is fed to the GPU for further rendering operations. As such, resource allocation, synchronization and buffer management of the decoding resources are critical in order to consume the content and meet the tight deadlines of viewport rendering refresh of XR applications. VDI will enable the efficient management of the decoding instances and offer to the XR applications an interface that will hide the underlying complexity of this management.

SD targets the composition aspects of immersive applications. To enable XR experiences, a virtual scene must be constructed by assembling a number of assets of different types. Current scene descriptor formats do not efficiently support real-time media and/or scene graph updates and/or interaction and other features essential for immersive XR applications. The integration of MPEG media into existing scene description formats (e.g. gITF$^{TM}$) is specifically addressed by the MPEG-I SD specification.

## INTERFACES FOR VIDEO DECODING PLATFORMS

### The Conventional MPEG Video Decoding Model

MPEG has published standards that have been a formidable accelerator for the mass distribution of digitally encoded media. MPEG standards range from compression, storage and to more recently streaming formats such as MPEG-DASH. In April 1996, ISO published the first edition of ISO/IEC 13818 part 1 Systems (1), a.k.a. MPEG-2 Systems, and the seventh edition in June 2019. This part of MPEG-2 defines the model for processing and decoding audio and video coded representations and their presentation in a synchronised fashion. To this end, MPEG experts defined the System Target Decoder (STD) which is "a hypothetical reference model of a decoding process", as shown in Figure 1 for the STD program stream variant.

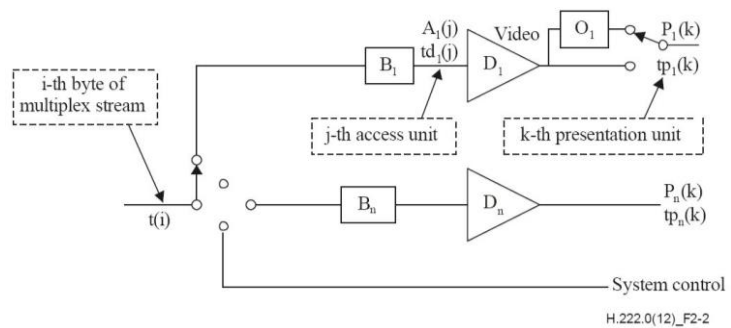

Figure 1 – Program stream System Target Decoder (P-STD), from ISO/IEC 13818-1

As seen in Figure 1, the i-th byte of the input program stream arrives at time t(i). Based on the elementary stream index n, the i-th byte is oriented to the pipeline associated with its elementary stream. For each elementary stream n, a decoder $D_n$ receives and decodes the j-th access unit $A_n(j)$ at time $td_n(j)$. The decoded access unit constitutes the k-th presentation unit $P_n(k)$ which is then presented at time $tp_n(k)$.

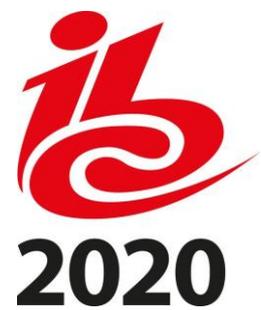

By defining the STDs (for both transport streams and program streams), a conventional decoding model (CDM) has been established and was successfully used for several decades. The rules governing this CDM are as follow:

i. Each elementary stream is decoded by one decoder instance, and thus:

    a. Each video access unit has a given picture resolution

    b. Each video access unit has a given chroma sampling

    c. Each elementary stream has a given (variable) frame rate

ii. Each decoded access unit constitutes one presentation unit

iii. Each access unit is associated with a presentation timestamp via the corresponding presentation unit referring to this access unit

**New Video Decoding Needs of Immersive Streaming Applications**

**Viewport-dependent streaming**
Streaming immersive applications are by essence event-driven and thus consume the media representations in a dynamic fashion. These events can be the user's head movement, the user's body translation, but also more traditional events for streaming applications such as bandwidth variations. All these frequent events are factored in by the immersive applications when retrieving the media content. The goal for the application is to maximise the quality of experience perceived by the user. Since immersive applications render the view based on the user viewport, the media content should also offer temporal and spatial random access. This way, the application can retrieve the piece of data it needs for a particular viewport at a particular time in the quality that is appropriate. In other words, immersive applications are performing viewport-dependent rendering and thus there is an ongoing trend to design viewport-dependent streaming logic such that the minimum amount of data is retrieved by the application for the maximum quality perceived by the user. An example of such a system is proposed by Graf et al (2) and reproduced in Figure 2 for streaming 360-degree videos using tiles formatted and described according to MPEG-DASH and the SRD extension (3).

This flexibility in terms of what is retrieved means that the elementary streams that come out of the encoders are generally not the input of the decoders in the application. Such elementary streams are often manipulated and these bitstream manipulations should be as lightweight as possible whether they are performed by media packagers, network entities or by the application itself.

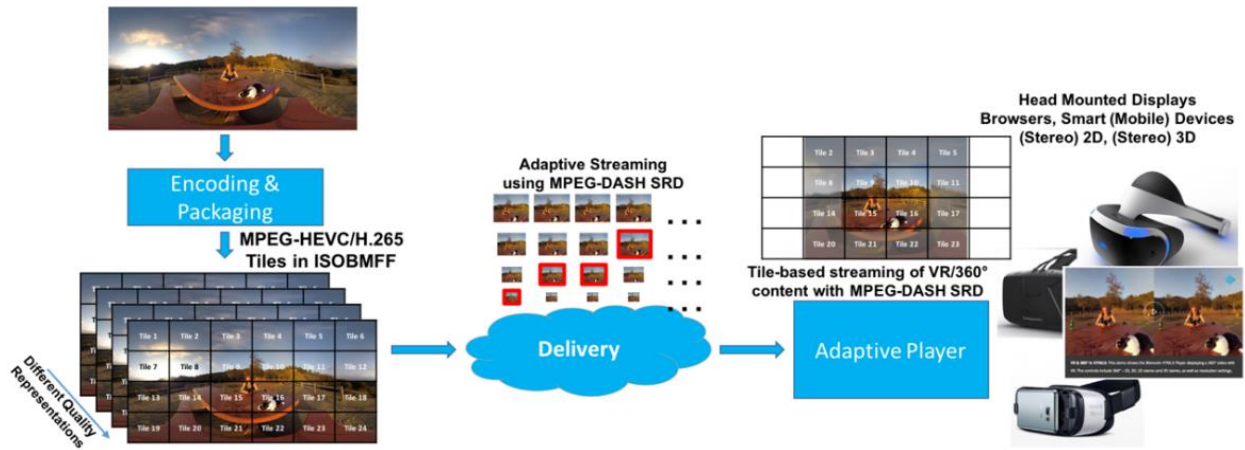

Figure 2 – System Architecture for Bandwidth Efficient Tiled Streaming (2)

**Time alignment of independent video elementary streams after decoding**

In some cases, an immersive experience comprises several independent media elementary streams that all together form a visual object. For instance, MPEG has published the MPEG-I part 5: Video-based point cloud compression (V-PCC) (3) which defines several video components such as texture, geometry, occupancy, etc… which all together form the point cloud after reconstruction. While each of these video components is independently coded, there needs to be a synchronisation step, a time alignment operation, at the output of each decoding process before all the decoded pictures can be used as input of the reconstruction process of the point cloud. Figure 3 shows the different video components constituting the coded point cloud as well the different input decoded sequences of the reconstruction operations.

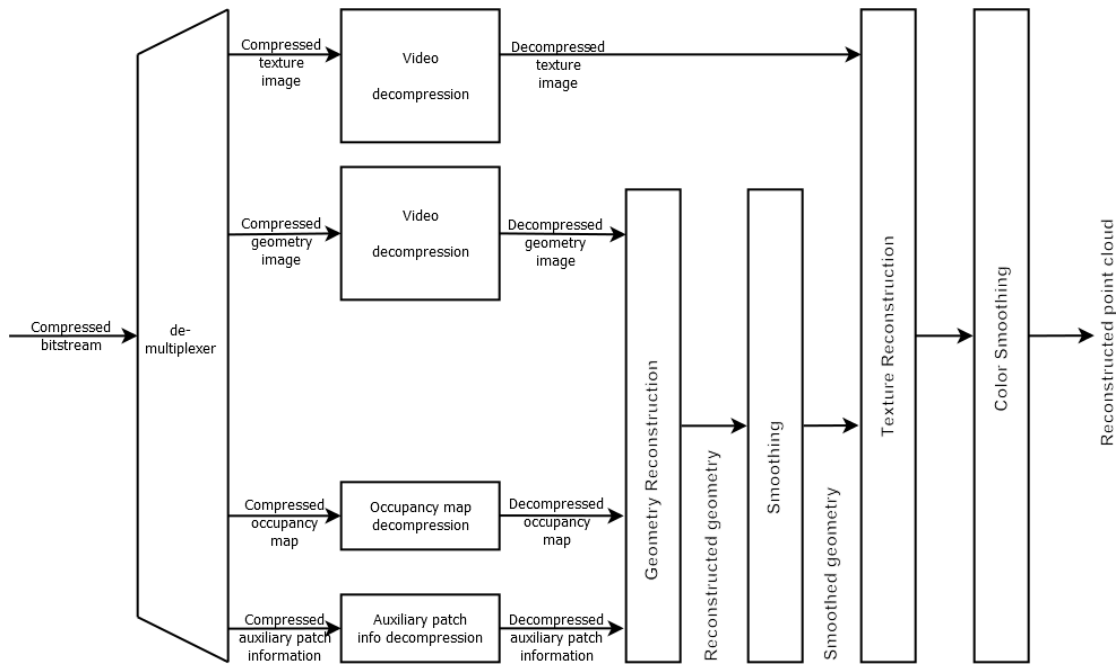

Figure 3 – V-PCC decoding structure (3)

While running several decoder instances on a video decoding platform is possible to the extent the hardware can actually support it, there is no guarantee that these parallel decoding instances would run in a synchronised fashion. On the contrary, implementing V-PCC demonstrators in MPEG showed that some of the decoder instances run ahead or behind of one another by several pictures. For example, the decoded texture picture of time t is used with the decoded geometry picture of time t+dt. This slight desynchrony introduces visual artefacts when performing the reconstruction operations of the point cloud with various levels of severity depending on how large the desynchrony is and which components are impacted.

Traditionally, the synchronisation of two video elementary streams is performed by the presentation engine based on timestamps. Here, the need arises for the immersive application to synchronise decoded pictures right after decoding based on their position in the stream (also known as Picture Order Count in some video coding standards) and before the presentation step at which the decoded data may be further synchronised based on timestamps in the conventional way. These two types of synchronisation are not only complementary but may also be sequentially needed given the type of application.

**MPEG-I Video Decoding Interface (VDI)**

**Scope**

MPEG-I VDI is the part 13 of MPEG-I published under the number ISO/IEC 23090-13 (5). The aim of VDI is to address the problems and challenges when implementing immersive applications as described in the paragraph "New Video Decoding Needs of Immersive Streaming Applications". To this end, the scope of the VDI specification covers the interface between a media application and the Video Decoding Engine (VDE) sitting on the device as shown in Figure 4.

A VDE as defined in VDI is the generic term corresponding to the application programming interface (API) exposing the capacity of the video decoding hardware platform of the device. Examples of VDIs are Khronos OpenMAX™ and the MediaSource Object of the Media

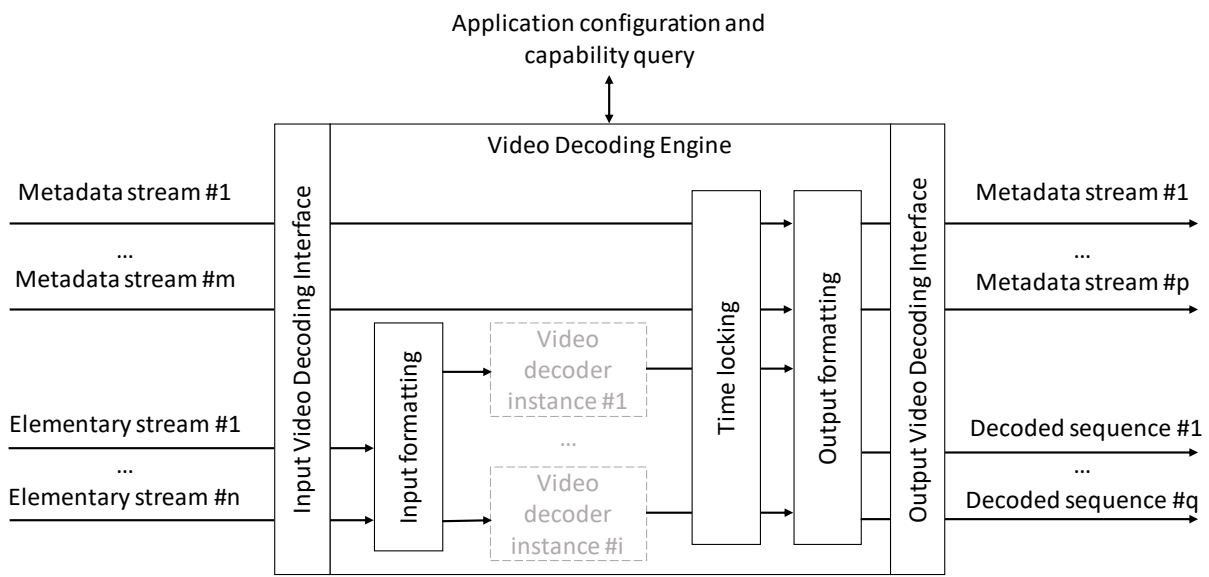

Figure 4 – Video Decoding Engine and interfaces

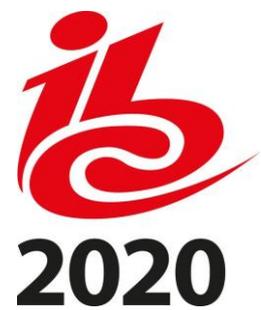

Source Extensions W3C Recommendation (6). From the same VDE, several decoder instances can be initiated. The purpose of the VDI specification is to provide to the application a certain level of orchestration of these concurrently running decoding instances via a set of functions, which are summarised in the paragraph "Operations and interfaces of the VDI".

**Operations and interfaces of the VDI**

The MPEG-I VDI specification defines new functions and operations on top of the existing ones provided by the underlying VDI, e.g. Khronos OpenMAX™. The list of new functions defined by the VDI specification as of May 2020 is provided in Table 1. The specification is still under development and additional functions are expected to be added and refined before the final publication of the specification. The main benefit of these new functions is that the application can not only spawn new decoding instances but can also group them into a group instance. This group instance would share common properties and will be coordinated in such a way that the several instances share the resources of VDE in a fair manner and in the interest of the application and not compete against each other.

| Function | Operation |
|---|---|
| queryCurrent Aggregate Capabilities | Query the instantaneous aggregate capabilities of a decoder platform for a specific codec component. |
| getInstance | Initiate a new decoding instance and optionally assign it to a group of existing instances. |
| setConfig | Configure a given instance in terms of output buffer properties |
| getParameter | Get dynamic parameters controlling a given instance, e.g. cropping window of the decoded pictures |
| setParameter | Set dynamic parameters controlling a given instance e.g. cropping window of the decoded pictures |

Table 1 – Description of the VDI functions

For instance, it should be avoided that one decoding instance would run ahead of another in terms of number of pictures processed; which can cause visual artefacts in the final rendering as presented earlier in this article.

In addition to these functions, new operations are also defined. As shown in Figure 4, the different types of operations are input formatting, time locking and output formatting. The draft specification as of May 2020 comprises four operations in the input formatting category, namely filtering, inserting, appending and stacking. As explained in the paragraph "New Video Decoding Needs of Immersive Streaming Applications", it is fairly common for immersive applications to produce content in parts, sometimes also called tiles. These tiles are independently encoded and share the same encoding parameters. As a result, these tiles can be indifferently decoded together as one video bitstream or in several video bitstreams, one for each tile. However, merging them into a single video bitstream is a tedious task for an application to perform since it requires rewriting and parsing of low-level data in the video bitstreams. The level of complexity of this task varies based on the video coding standard used for encoding the video elementary streams. Advantageously, the input formatting module will allow the VDI to achieve this operation in lieu of the application. This way, the number of incoming video elementary streams and the number of video decoder instances to initiate can be decoupled and no longer follows a one-to-one mapping as described in paragraph "The Conventional MPEG Video Decoding Model". Furthermore, the number of instances can be based on runtime optimisation decision and/or device capability discovery. For instance, a VDE may be able to decode one 4K video bitstream but not necessarily 4 HD video bitstreams simultaneously in which case merging them, if permitted by the bitstreams constraints, would allow their decoding. As hinted in this description, the input formatting module does expect certain constraints on the elementary streams, namely the video codec used and some constraints on the input video bitstreams. For that purpose, the VDI specification defines binding of the four input formatting operations with specific video codecs, namely with the Versatile Video Codec Coding (VVC) / H.266 (7) and expectedly with the High Efficiency Video Coding / H.265 (8) specifications. More codec bindings may be later added during the standardisation phase of the specification.

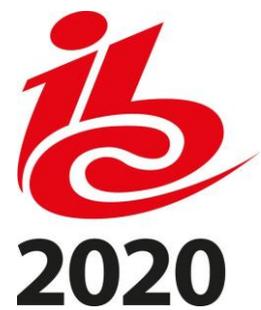

Along with the specification, a sample library and a conformance software are being developed to validate the proper integration of the different modules of the VDI as well as the definition and constraints on the codec bindings. The publication of the VDI specification and the accompanying software is scheduled for 2021.

## SCENE DESCRIPTION

### Introduction to Scene Description

Another key technology for enabling immersive 3D user experiences is scene description. Scene description is used to describe the composition of a 3D scene, referencing and positioning the different 2D and 3D assets in the scene. The information provided in the scene description is then used by a presentation engine to render the 3D scene properly, using techniques like Physically-Based Rendering (PBR) that produce realistic scenes.

A scene description usually comprises a scene graph which is a directed acyclic graph, typically a plain tree-structure, that represents an object-based hierarchy of the geometry of a scene. The leaf nodes of the graph represent geometric primitives such as images, textures or media data buffers. Each node in the graph holds pointers to its children. The child nodes can among others be a group of other nodes, a geometry element, a transformation matrix, accessors to media data buffers, camera information for the rendering, etc.

Spatial transformations are represented as nodes of the graph and represented by a transformation matrix. Typical usage of transform nodes is to describe rotation, translation or scaling of the objects in its child nodes. Scene graph also supports animation nodes that allow changes to animation properties over time, hence describing dynamic content.

This structure of scene graphs has the advantage of reduced processing complexity, e.g. while traversing the graph for rendering. An example operation, that is simplified by the graph representation, is the culling operation where branches of the graph are dropped from processing, if deemed that the parent node's space is not visible or relevant (level of detail culling) to the rendering of the current view frustum.

While there are many proprietary solutions for scene description (typically at the heart of game engines, VFX design tools or AR/VR authoring tools), several solutions have also been standardized. In particular, Virtual Reality Modeling Language (VRML), which uses XML syntax, was the first scene description solution to be standardized in 2001 for WEB usages. Later on, OpenSceneGraph, an open source project using OpenGL which was released in 2005, has been used as a component in several computer games and rendering platforms such as Delta3D or FlightGear. In 2010, X3D, standardized by ISO/IEC, became the successor of VRML, introduced binary formats and JSON format for scene graph description and featured new capabilities such as multi-texture rendering, shading, real-time environment lightning, and culling. Finally, more recently in 2015, the Khronos Group, who also manages OpenGL, released the Graphics Library Transmission Format (glTF). glTF is a scene description format, based on a JSON format, intended to be efficient and interoperable as a common description format for 3D content tools and services.

For its own scene description approach, MPEG decided to base its work on glTF as this technology is already widely deployed and includes an extension mechanism. By working on the definition of such extensions, MPEG can precisely focus on the shortcomings of the current glTF solution with respect to MPEG media support and proposes its own solution as an MPEG branded glTF extension.

A typical glTF scene graph is a tree of nodes describing content properties, access to content data, possible dynamic animations and parameters for the rendering whose relationship are shown in Figure 5. Typically, a camera node describes the view from which rendering shall be made, 3D objects are described in mesh sub-nodes whose child nodes provide information on how to access the media data through buffers and texture information, and animation nodes describe dynamic changes of 3D objects over time.

**MPEG Extensions to glTF 2.0**

glTF 2.0 (9) provides a solid and efficient baseline for exchangeable and interoperable scene descriptions. The conceptual relationship between the top-level elements of a glTF asset is given in Figure 5. Traditionally, glTF 2.0 has been focused on static scenes made of static assets, which makes it unfit to address the requirements and needs of dynamic and rich 3D scenes in immersive environments.

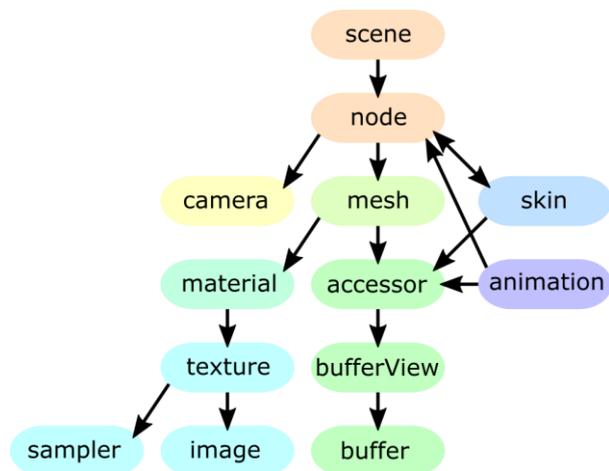

Figure 5 – Conceptual relationship between the top-level elements in glTF (9)

As part of its effort to define solutions for immersive multimedia, MPEG has identified the following gaps in glTF 2.0:

- No support for timed media like video and moving meshes and point clouds.
- No support for audio.
- Limited support for interactions with the scene and the assets in the scene.
- No support for local and real-time media, which are crucial for example for AR experiences.

MPEG has decided to leverage the extension mechanism that glTF 2.0 offers, to develop its solution for immersive multimedia after carefully evaluating different alternatives, including home grown old scene graph solutions and the possibility of defining a new one from scratch.

In this section, we give a brief overview of the extensions that have been developed by MPEG so far to address the identified gaps. MPEG developed an architecture for MPEG-I to guide the work on the scene description, which serves as the entry point for consumption of an immersive media presentation. Figure 6 depicts the MPEG-I architecture and defines the key interfaces.

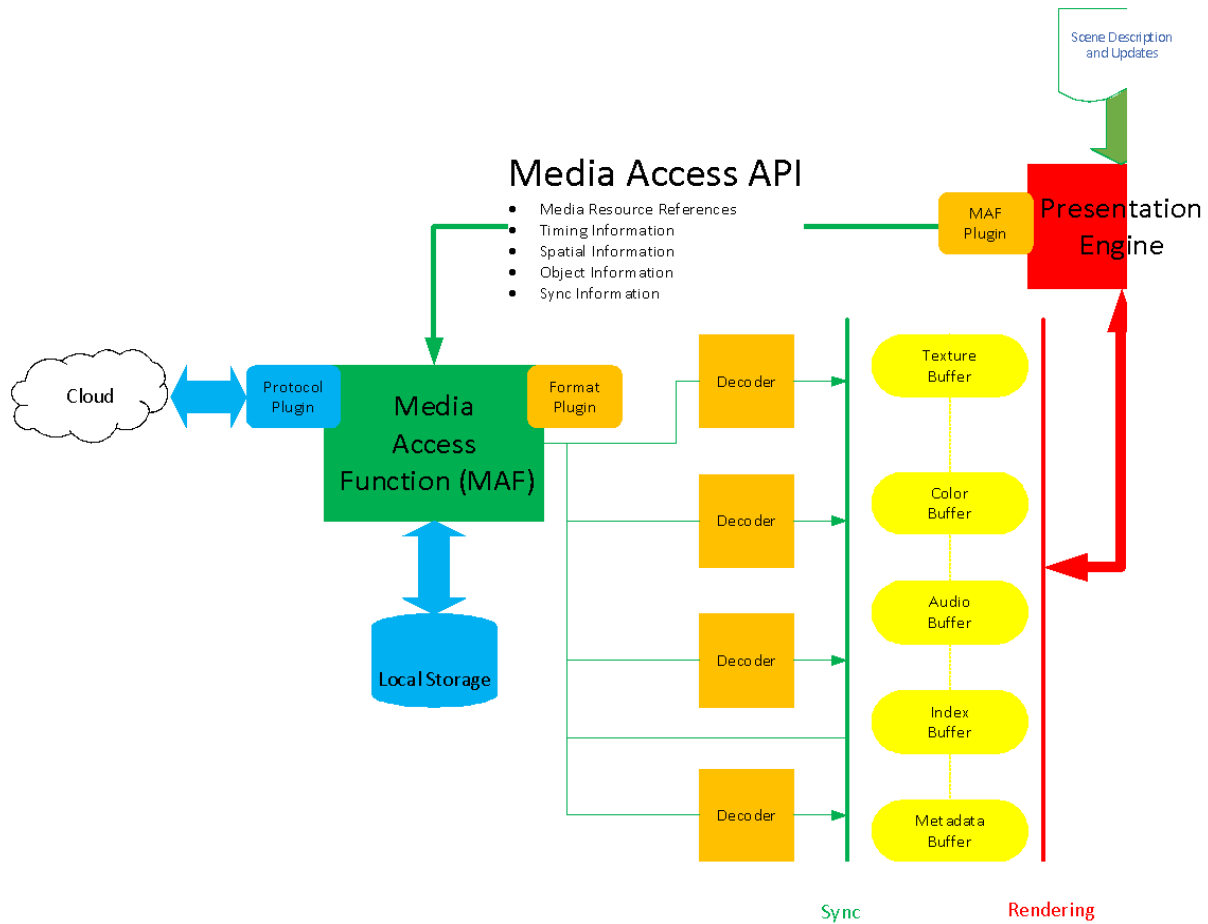

Figure 6 – Scene Description Reference Architecture

The design focuses mainly on buffers as means for data exchange throughout the media access and rendering pipeline. It also defines a Media Access API to request media that is referenced by the scene description, which will be made accessible through buffers. This design aligns with glTF 2.0 principles and integrates with VDI presented earlier in this paper.

**MPEG_timed_accessors extension**
In order to provide access to timed media and metadata in a scene, a new glTF extension is specified to define timed accessors. An accessor in glTF defines the types and layout of the data as it is stored in a buffer that is viewed through a `bufferView`.

When timed data is read from a buffer, the data in the buffer is expected to change dynamically with time. The buffer element is extended to add support for a circular buffer that is used with timed data.

## MPEG_circular_buffer extension

The glTF 2.0 buffer element is extended to provide circular buffer functionality. The extension is named `MPEG_circular_buffer` and may be included as part of the `buffer` structures. Buffers that provide access to timed data must include the `MPEG_circular_buffer` extension.

Frames of the buffer may differ in length based on the amount of data for each frame. A read and a write pointer are maintained for each circular buffer. By default, read and write access to the buffer will be served from the frame that is referenced by the read or write pointer respectively. Access to a particular frame index or timestamp should be supported.

The frames are read at the read pointer for rendering. New incoming frames from the media decoder are inserted at the write pointer. Prior data in that frame will be overwritten and the frame buffer should be resized accordingly.

The renderer will always maintain that read operations are performed on stored data and in an asynchronous way to the write operations to the buffer. This will ensure that no deadlock situations arise from simultaneous write operations by the Media Access Function and the Presentation Engine.

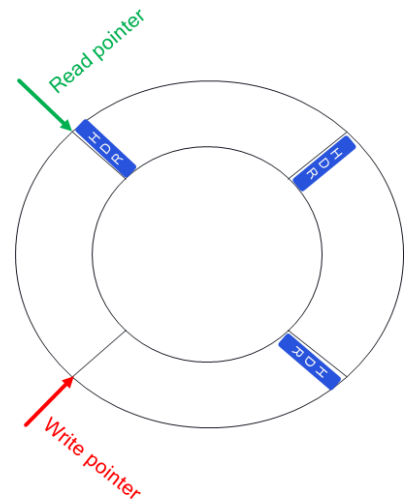

Figure 7 – Buffer structure

## MPEG_media extension and MPEG_video_texture

The MPEG media extension, identified by `MPEG_media`, provides an array of media items used by different assets in the scene. This extension provides the necessary information to make requests through the MAF API for media data.

MPEG video texture extension, identified by `MPEG_video_texture`, provides the possibility to link a glTF texture object to a media and its respective track listed by `MPEG_media` object. The MPEG video texture extension references a timed accessor, using the `timedAccessor` object to receive the decoded timed texture for rendering in form of a circular buffer.

## MPEG audio extension

The MPEG audio extension adds support for spatialized audio to the MPEG scene description based on glTF 2.0. This extension is identified by `MPEG_spatial_audio`, which can be included at top level or attached to any node in the scene.

The MPEG_spatial_audio extension supports four different node types:

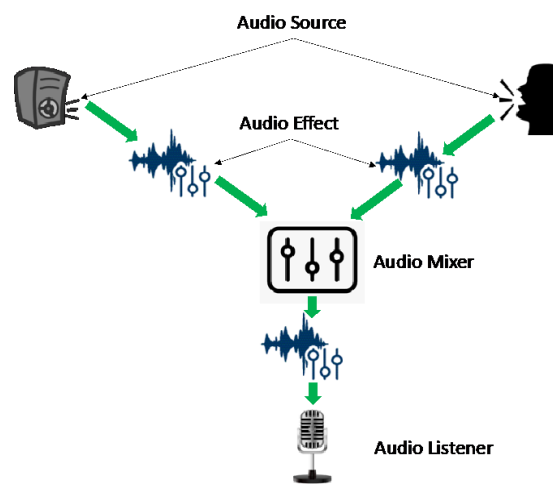

Figure 8 – Processing chain for audio in a scene

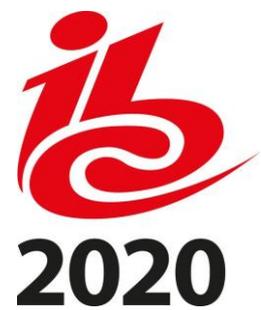

- `AudioSource`: an audio source that provides input audio data into the scene
- `AudioMixer`: an audio mixer that mixes the output of one or more audio sources, effects, and other mixers to produce an output audio signal.
- `AudioEffect`: An effect can be attached to the output of a source or a mixer or to the input of an audio listener. It may also be standalone, in which case, it will apply to all audio listeners that are in their active zone in the scene.
- `AudioListener`: An audio listener represents the output of audio in the scene. They are usually attached to camera nodes in the scene.

The characteristics of `AudioListener` depend on the actual output devices available to the audio renderer.

**Scene Updates**

In addition to the listed extensions, a Scene Update mechanism has been developed by MPEG. The Scene Updates are expressed using the JSON Patch protocol as defined in RFC 6902. Each update operation consists of a JSON Patch document, where all update operations are considered as a single timed transaction. All update operations of a transaction are executed successfully for an update operation to be considered successful.

After successfully performing an update operation, the resulting scene graph must remain consistent, valid, and all references must be correct. Since glTF 2.0 uses the order of elements for referencing, particular care is taken with update operations that change the order of elements in the graph, such as move and remove operations. The client must update all references after every successful scene update operation.

**CONCLUSION**

MPEG-I VDI and SD provide enablers for richer XR applications using MPEG media. The combination of both specifications is expected to lower the barrier in terms of bandwidth and latency requirements for streaming high quality XR experiences. Among other things, the viewport-dependent streaming approach will be facilitated for XR applications by integrating encapsulated MPEG media into existing scene description formats as well as enhancing the video decoding platform APIs with interfaces to facilitate the management of multiple concurrent video decoding instances. Both new MPEG specifications will offer software conformance, test vectors and sample libraries for validating every step of the process of properly integrating with the existing XR ecosystem. The publication of both specifications is expected for end of 2021.

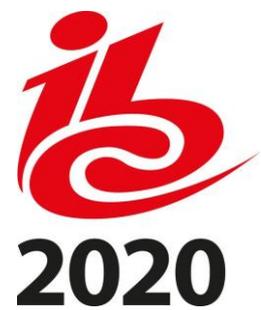